\documentclass[prd,twocolumn,showpacs,amsfonts,amssymb,amsmath]{revtex4-1}

\usepackage{graphicx}
\usepackage{simplewick}
\usepackage{epstopdf}
\usepackage{multirow}
\usepackage{dcolumn}
\usepackage{hyperref}
\usepackage{cases}
\usepackage{bm}
\usepackage{epsfig}
\usepackage{color}
\usepackage[normalem]{ulem}

\def\be{\begin{equation}}
\def\ee{\end{equation}}
\def\bea{\begin{eqnarray}}
\def\eea{\end{eqnarray}}

\newcommand{\bear}{\begin{eqnarray}}

\newcommand{\eear}{\end{eqnarray}}

\newlength{\tskip}
\newbox\pippobox

\def\be{\begin{equation}}
\def\ee{\end{equation}}
\def\bea{\begin{eqnarray}}
\def\eea{\end{eqnarray}}

\def\9{\nabla}

\def\6{\partial}

\def\f{\frac}

\def\0{(0)}

\def\>{\rightarrow}

\begin{document}

\title{{\bf EFTCAMB/EFTCosmoMC: massive neutrinos in dark cosmologies}}

\author{Bin Hu$^1$, Marco Raveri$^{2,3}$, Alessandra Silvestri$^{1}$, Noemi Frusciante$^{2,3}$}
\affiliation{
\smallskip
$^{1}$ Institute Lorentz, Leiden University, PO Box 9506, Leiden 2300 RA, The Netherlands \\
\smallskip 
$^{2}$ SISSA - International School for Advanced Studies, Via Bonomea 265, 34136, Trieste, Italy \\
\smallskip
$^{3}$ INFN, Sezione di Trieste, Via Valerio 2, I-34127 Trieste, Italy}

\begin{abstract}
We revisit the degeneracy between massive neutrinos and generalized theories of gravity in the framework of  effective field theory of cosmic acceleration. In particular we consider $f(R)$ theories and a class of non-minimally coupled models parametrized via a coupling to gravity which is linear in the scale factor. In the former case, we find a slightly lower degeneracy with respect to what found so far in the literature, due to the fact that we implement exact designer $f(R)$ models and evolve the full linear dynamics of perturbations. As a consequence, our bounds are  slightly tighter on the $f(R)$ parameter but looser on the summed neutrino mass. We also set a new upper bound on the Compton wavelength parameter ${\rm Log}_{10}B_0<-4.1$ at $95\%$ C.L. with fixed summed neutrino mass ($\sum m_{\nu}=0.06$ eV) in $f(R)$ gravity with the combined data sets from cosmic microwave background temperature and lensing power spectra of {\it Planck} collaboration as well as galaxy power spectrum from WiggleZ dark energy survey.  We do not observe a sizable degeneracy between massive neutrinos and modified gravity in the linear parametrization of non-minimally gravitational coupling model. The analysis is performed with an updated version  of the EFTCAMB/EFTCosmoMC package, which is now publicly available and extends the first version of the code with the consistent inclusion of massive neutrinos, tensor modes, several alternative background histories and designer quintessence models. 
\end{abstract}


\maketitle

%
\section{Introduction}\label{Sec:Intro}
The direct measurements of neutrino flavor oscillations provide evidence for non-zero neutrino masses, but give no hint on their absolute mass scale~(see e.g. the reviews~\cite{Maltoni:2004ei,Fogli:2005cq}). Cosmology, on the other hand, provides a powerful, complementary way of placing constraints on the sum of the mass of neutrinos~(see e.g. the reviews~\cite{Lesgourgues:2006nd,Wong:2011ip}). Indeed massive neutrinos can significantly affect the distribution of large-scale structure (LSS) and the pattern of cosmic microwave background (CMB) anisotropies depending on the value of their mass. The current constraint from CMB experiments on the summed neutrino mass fixes the upper limit at $\sum{m_{\nu}}<0.66$ eV (95\%; {\it Planck}+WP+highL) for a flat $\Lambda$CDM cosmology~\cite{Ade:2013zuv}. Besides slightly affecting the expansion history, massive neutrinos leave an imprint on the dynamics of linear scalar perturbations. On scales smaller than their mass scale neutrinos free stream damping the structure  and accordingly  diminishing the weak lensing effect on those scales~\cite{Lewis:2002nc}. Furthermore, they contribute an early Integrated-Sachs Wolfe (ISW) effect because their transition from relativistic to non-relativistic happens on an extended redshift interval which, for the typical neutrino mass ($\sum m_{\nu}\sim 0.1$ eV), overlaps with the transition from radiation to matter~\cite{Lewis:2002nc}.

Similar effects are also observed in dark energy (DE) and modified gravity (MG) models that address the phenomenon of  cosmic acceleration. The latter generally involve an extra, dynamical massive scalar d.o.f. which mediate a fifth force between matter particles and can have a speed of sound different from unity. Besides affecting the background dynamics such as to source cosmic acceleration, on linear scales the field can modify significantly the clustering of matter as well as the sub-horizon dynamics of metric potentials on scales below or above its characteristic lengthscale. Hence, structure formation, the ISW effect and weak lensing effect will be modified accordingly~\cite{Silvestri:2009hh,Jain:2010ka,Jain:2013wgs}.
Based on these considerations, a degeneracy between massive neutrinos and the dark sector is expected, and in general neutrinos bounds depend significantly on the cosmological model within which they are analyzed.  This  has been investigated by several authors~\cite{Hojjati:2011ix, Motohashi:2012wc, Baldi:2013iza, He:2013qha, Dossett:2014oia}.  

In this paper we investigate the degeneracy between massive neutrinos and cosmological models which deviate from general relativity  by the inclusion of an extra scalar degree of freedom in the framework of effective field theory (EFT). In particular, we extensively analyze $f(R)$ theories, with the designer approach, updating the bound on the Compton scale parameter $B_0$ with and without massive neutrino. We also consider a linear EFT parametrization of non-minimally coupled models. We use mainly  CMB and LSS observables as described in detail in Section~\ref{data}. The present investigation is made by means of EFTCAMB and EFTCosmoMC~\cite{Hu:2013twa,Raveri:2014cka}. These are patches of CAMB/CosmoMC~\cite{CAMB,Lewis:1999bs,Lewis:2002ah} which allow to investigate the evolution of linear perturbations in a model independent way as well as in any specific DE/MG model that can be cast into EFT framework of cosmic acceleration formulated by~\cite{Gubitosi:2012hu,Bloomfield:2012ff,Piazza:2013coa}. 

A new release of EFTCAMB which is fully compatible with massive neutrinos is now available at \url{http://wwwhome.lorentz.leidenuniv.nl/~hu/codes/}. Let us note that the new version of EFTCAMB/EFTCosmoMC has been also equipped with  several alternative DE equation of state parametrizations~\cite{Jassal:2004ej,Jassal:2006gf,Hu:2014ega}, the tensor perturbations equation and  designer minimally coupled quintessence models. This release comes with  detailed notes~\cite{Hu:2014oga}.  
\section{Observables and Data}\label{data}
In our analysis we  will use different combinations of the following data sets. 
We employ the {\it Planck} temperature-temperature power spectra considering the 9 frequency channels ranging from $30\sim353$ GHz for low-$\ell$ modes ($2\leq\ell<50$) and the $100$, $143$, and $217$ GHz frequency channels  for high-$\ell$ modes ($50\leq\ell\leq2500$)~\cite{Ade:2013kta,Ade:2013zuv}.
In addition we include the WMAP low-$\ell$ polarization spectra ($2\leq\ell\leq32$)~\cite{Hinshaw:2012aka} in order to break the degeneracy between the re-ionization optical depth and the amplitude of CMB temperature anisotropy.
In the following we will denote the combination of the two above data sets as PLC. 
We also include the {\it Planck} 2013 full-sky lensing potential map~\cite{Ade:2013tyw}, obtained by using the $100$, $143$, and $217$ GHz frequency bands that resulted in a detection of the CMB lensing signal with a significance greater than $25\sigma$. We will refer to this data set as the lensing one. 
The baryon acoustic oscillations measurements, that we will denote as BAO, are taken from the 6dFGS ($z= 0.1$)~\cite{Beutler:2011hx}, SDSS DR7 (at effective redshift $z_{\rm eff}=0.35$)~\cite{Percival:2009xn,Padmanabhan:2012hf}, 
and BOSS DR9 ($z_{\rm eff}=0.2$ and $z_{\rm eff}=0.35$)~\cite{Anderson:2012sa} surveys. 
Finally we use measurements of the galaxy power spectrum as made by the WiggleZ Dark Energy Survey~\cite{wigz} in order to exploit the constraining power of data from large-scale structure. 
This latter data set consists of the galaxy power spectrum measured from spectroscopic redshifts of $170,352$ blue emission line galaxies over a volume of $1\,\mbox{Gpc}^3$~\cite{Drinkwater:2009sd,Parkinson:2012vd} and the covariance matrices as given in~\cite{Parkinson:2012vd} are computed using the method described by~\cite{Blake:2010xz}.
It has been shown that linear theory predictions are a good fit to the data regardless of non-linear corrections up to a scale of $k\sim 0.2\,\mbox{h}/\mbox{Mpc}$~\cite{Parkinson:2012vd,Dossett:2014oia} and for this reason, in this work, we use the WiggleZ galaxy power spectrum with $k_{\rm max} = 0.2\,\mbox{h}/\mbox{Mpc}$.
Finally,  we marginalize over a linear galaxy bias for each of the four redshift bins, as in~\cite{Parkinson:2012vd}.
\section{A worked example I: massive neutrinos and $f(R)$ models}\label{Sec:FRmassiveNu}
As we discussed in the Introduction, massive neutrinos are an extension of the cosmological standard model which modify the dynamics of linear scalar perturbations, leaving a characteristic imprint on the growth of structure~\cite{Lesgourgues:2006nd}. Specifically, on linear scales smaller than the neutrino free streaming distance, the overall matter clustering is suppressed. Interestingly scalar tensor models of modified gravity leave a complementary signature on the growth of structure, enhancing the clustering on linear scales within the Compton scale of the extra scalar degree of freedom, because of the fifth force mediated by the latter. Depending on the mass of neutrinos and of the scalar field, there may be a significant degeneracy between the two effects at some redshifts and scales. The latter has been investigated to large extent in the context of $f(R)$ theories of gravity, and generally an appreciable degeneracy has been found. However, a common feature of all previous analyses is the 
assumption of the quasi static (QS) limit in the equations for the perturbations and the employment of the QS parametrization introduced by Bertschinger and Zukin in~\cite{Bertschinger:2008zb}. In this paper we revisit this degeneracy employing EFTCAMB, which has the important virtue of letting us implement exact $f(R)$ models and evolve their full dynamics.  Another key feature of our analyses is the consistent treatment of the background cosmology, which is based on a designer reconstruction of $f(R)$ models with the inclusion of massive neutrinos. 
\subsection{Numerical implementation of $f(R)$ models}\label{Sec:IIA}
We shall now discuss in some detail the implementation of $f(R)$ models into EFTCAMB. We will then review their implementation in MGCAMB~\cite{Zhao:2008bn,Hojjati:2011ix}, since in the following subsection we will compare the results obtained with these two codes.

$f(R)$ gravity is described by the following action in Jordan frame
\be\label{action_fR}
S=\f{M_P^2}{2}\int{} d^4x \sqrt{-g} \left[R+f(R)\right]+S_m\,,
\ee
where $f(R)$ is a  general function of the Ricci scalar R, and  $S_m$ is the  matter action whose Lagrangian is minimally coupled to gravity. For a detailed review of the cosmology of $f(R)$ theories we refer the reader to~\cite{Sotiriou:2008rp,DeFelice:2010aj,Lombriser:2014dua}. Despite the fourth order nature of the $f(R)$ field equations, it can be shown that the above action belongs to the class of scalar tensor theory with second order field equations, where the role of the scalar d.o.f. is played by $f_R\equiv df/dR$, commonly dubbed the scalaron~\cite{Starobinsky:2007hu}. The dynamics of linear scalar perturbations in models of $f(R)$ gravity has been extensively studied in~\cite{Song:2006ej,Bean:2006up,Pogosian:2007sw}. Here, we shall focus on the designer approach to $f(R)$ theories that EFTCAMB exploits. It has been introduced in~\cite{Song:2006ej}  and offers a practical way of reconstructing all $f(R)$ models that reproduce a given expansion history. Once the latter is chosen, the modified Friedmann equation can be solved as a second order differential equation for $f[R(a)]$. While the original setup considered only dust in the energy budget of the Universe, in~\cite{Pogosian:2007sw} the approach was extended to include radiation and in~\cite{He:2013qha} to include also massive neutrinos via the approximation adopted by WMAP7~\cite{Komatsu:2010fb}. Here, we consistently include the massless/massive neutrino sector into the designer formalism by using instead the formula adopted by CAMB~\cite{Lewis:2002nc,Howlett:2012mh}. 

Let us define the following dimensionless quantities
\be
\label{Eq:DesignerDefinitions}
y\equiv   \frac{f(R)}{H_0^2}, \hspace{0.3cm}E \equiv  \frac{H^2}{H_0^2} =E_{\rm m} +E_{\rm r}+E_{\nu}+E_{\rm{eff}}.  
\ee
with:
\begin{align}
\label{Eq:EnergySpecies}
&E_{\rm m} = \Omega_{\rm m} a^{-3} \;,\hspace{0.3cm}
E_{r}  = \Omega_{r} a^{-4} \;, \hspace{0.3cm}
E_{\nu} =\frac{8\pi G}{3H_0^2}\rho_{\nu} \;,\nonumber \\
&E_{\rm{eff}} = \Omega_{\rm eff} \exp\left[-3\ln a+3\int_a^1 w_{\rm{eff}}(\tilde{a}) d\ln\tilde{a} \right],
\end{align}
where the modifications of gravity are treated as an effective dark fluid  with an equation of state $w_{\rm{eff}}(a)$ and where we have enforced flatness  $\Omega_{\rm eff}=1-\Omega_m-\Omega_r-\Omega_{\nu}$. 
To be clear, here the subscripts ``m'', ``r'' and ``$\nu$'' denote, respectively,  non-relativistic matter (baryons and CDM), photons and neutrinos (including both massless and massive species). 
We assume that all massive neutrino species have equal masses and that neutrino decoupling in the early universe is instantaneous~\cite{Lewis:2002nc} so that their distribution function is given by the Fermi-Dirac distribution. 

When neutrinos are in the relativistic regime their momentum is large compared to their rest mass $q\gg m_{\nu}a$ and their energy density and pressure are given by:
\begin{align}
\label{Eq:RelativisticNeutrinoDensityPressure}
\rho_{\nu} =&\rho_{\nu (m=0)}\left(1+\frac{5}{7\pi^2}\bar m^2a^2\right) \;, \nonumber \\
P_{\nu} =&\frac{1}{3}\rho_{\nu (m=0)}\left(1-\frac{5}{7\pi^2}\bar m^2a^2\right)\;,
\end{align} 
where $\rho_{\nu (m=0)}=N_{\rm eff}7/8(4/11)^{4/3}\rho_{r}$ and $\bar m=m_{\nu}/k_BT_da_d$ are the massless neutrino density and dimensionless neutrino mass parameter. $N_{\rm eff}$ is the effective number of neutrino species whose standard value is $3.046$~\cite{Mangano:2001iu,Mangano:2005cc} and $T_d$ and $a_d$ are the temperature and scale factor at neutrino decoupling, respectively. 
In the non-relativistic regime the bulk of the neutrinos have $q\ll m_{\nu}a$ and their energy density and pressure evolve in the FRW limit as:
\begin{align}
\label{Eq:NonRelativisticNeutrinoDensityPressure}
\rho_{\nu} \simeq&\frac{180}{7\pi^4}\rho_{\nu (m=0)}\left(\zeta_3\bar ma+\frac{15\zeta_5}{2\bar ma}-\frac{945\zeta_7}{16(\bar ma)^3}+\cdots\right)\;, \nonumber \\
P_{\nu} \simeq&\frac{900}{7\pi^4}\rho_{\nu (m=0)}\left(\frac{\zeta_5}{\bar ma}-\frac{63}{4}\frac{\zeta_7}{(\bar ma)^3}+\cdots\right)\;,
\end{align}
where $\zeta_s$ are Riemann zeta functions.  
In the intermediate regime their density and pressure are given by the numerical integral of the Fermi-Dirac distribution.

As shown in~\cite{Song:2006ej}, the Friedmann equation for $f(R)$ theories can be written as a second order differential equation for $y$
\be\label{eq:designer}
y''-\left(1+\f{E'}{2E}+\f{R''}{R'}\right)y'+\f{R'}{6H_0^2E}y=-\f{R'}{H_0^2E}E_{\rm eff}\,,
\ee
where a prime indicates derivation w.r.t. $\ln a$ and $H_0$ is the present day Hubble parameter. As written explicitly in~(\ref{Eq:DesignerDefinitions}), $E$ contains also the contribution from neutrinos. 
We fix initial conditions deep in radiation dominated epoch ($a\sim10^{-8}$), when the effective dark component does not affect the evolution and the neutrinos are ultra-relativistic.  
The analysis in~\cite{Pogosian:2007sw} is still valid but the inclusion of massive neutrinos affects the physical processes after matter-radiation equality.  In particular their inclusion shifts the time of matter to radiation equality since at so early times neutrinos are in the ultra-relativistic regime. We emphasize that besides in the fitting formula of the neutrino sector, our work and~\cite{He:2013qha} differ also in the epoch when initial conditions of the designer approach are set up;  in~\cite{He:2013qha} the initial conditions are fixed  in the matter dominated era, while here they are set up during the radiation dominated epoch. 

After fixing the expansion history, we are ready to solve the designer equation~(\ref{eq:designer}). Following the argument in~\cite{Song:2006ej} we set to zero the amplitude of  the decaying mode at initial time and we are left with only one free boundary condition. In other words, for a given expansion history we will find a family of $f(R)$ models reproducing it and differing by the boundary condition. The latter is typically chosen to coincide with the present day value, $B_0$,  of the Compton wavelength in the Hubble unit
\begin{align}
B =\frac{2}{3(1+f_R)}\frac{1}{4 E'+E''}\frac{E}{E'}\left(y'' -y'\frac{4 E''+E'''}{4 E' + E''} \right)\,.
\end{align}
After an expansion history and a value for $B_0$ are chosen, EFTCAMB evolves the \emph{full}, linear, dynamics of perturbations  for the corresponding $f(R)$ model. 
\begin{figure*}[!th]
\begin{center}
\includegraphics[width=0.84\textwidth]{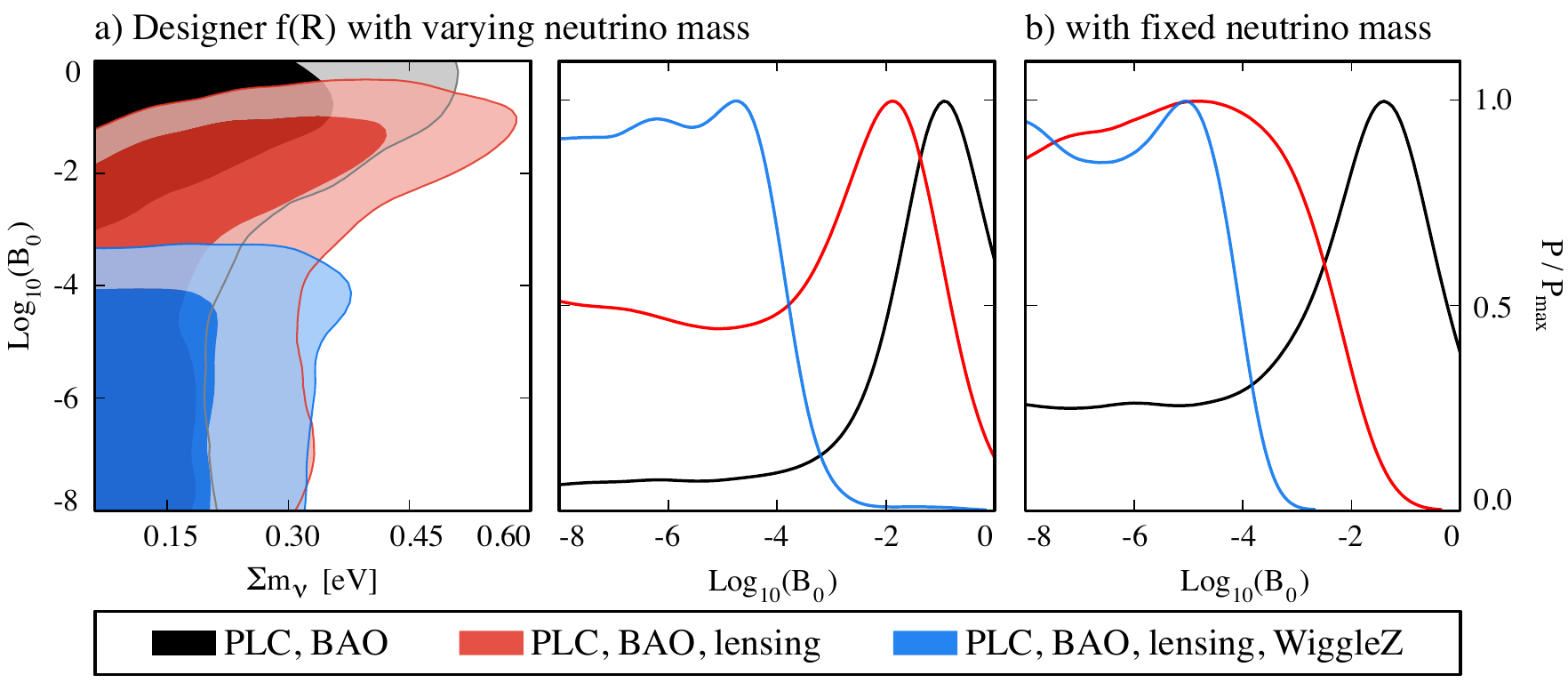}
\caption{{\it Left:} the marginalized joint likelihood for the present day value of  ${\rm Log}_{10} B_0$ and the sum of neutrino masses, $\sum m_\nu$. {\it Center and right:} the marginalized likelihood of  ${\rm Log}_{10} B_0$ for, respectively, designer $f(R)$ models with varying (a) and fixed (b) neutrino mass. In all three panels different colors correspond to different combination of cosmological observations as shown in legend. The darker and lighter shades correspond respectively to the $68\%$ C.L. and the $95\%$ C.L..}
\label{Fig:B0Experiment}
\end{center}
\end{figure*}
\setlength\tabcolsep{1pt}
\begin{table*}[htb!]
\footnotesize
\centering
\begin{tabular}{|l||c|c||c|}
\hline
\multicolumn{1}{|l||}{  }&
\multicolumn{2}{c||}{Varying $m_{\nu}$} &
\multicolumn{1}{c|}{Fixing $m_{\nu}$ } \\
\hline
\hline
Data sets & ${\rm Log}_{10} B_0$ $(95\% \,\,{\rm C.L.})$ & $\sum m_{\nu}$ $(95\% \,\,{\rm C.L.})$ & ${\rm Log}_{10} B_0$ $(95\% \,\,{\rm C.L.})$  \\
\hline
PLC + BAO 								  & none & $< 0.37$  & none \\
PLC + BAO + lensing					  & $<-1.0$ & $< 0.43$ &  $<-2.3$ \\
PLC + BAO + lensing + WiggleZ  & $< -3.8$ & $< 0.32$ &  $< -4.1$ \\
\hline
PLC + BAO + WiggleZ (EFTCAMB) & $< -3.8$ & $<0.30$ & 	 \\
PLC + BAO + WiggleZ (MGCAMB)  & $< -3.1$ & $<0.23$ &   \\
\hline
\end{tabular}
\caption{{\it Second column}: joint constraints on the $B_0$ parameter of designer $f(R)$ models and a varying neutrino mass, for different combinations of data sets as obtained with EFTCAMB; the last two rows report the comparison between EFTCAMB and MGCAMB for the specific combinations of data used in~\cite{Dossett:2014oia}. {\it Third column}: constraints on $B_0$ in the case of a fixed neutrino mass ($\sum m_\nu =0.06 {\rm eV}$) as obtained with EFTCAMB. In both cases, we use  different combinations of the data sets described in Section~\ref{data}.}
\label{Tab:ConstraintsFR}
\end{table*}
Things are different in the case when MGCAMB is used. While the framework at the basis of this code is in general not restricted to the quasi static limit, the implementation of specific models like $f(R)$ relies on the parametrization introduced by Bertschinger and Zukin in~\cite{Bertschinger:2008zb}, and later extended in~\cite{Pogosian:2007sw,Giannantonio:2009gi,Bean:2010zq}, which is quasi static and introduces an approximation for the time evolution of $f_R$. More specifically, the background is fixed to whatever the desired one is, in this case $\Lambda$CDM plus the parameter $f_\nu$ modeling massive neutrinos. 
The effects of modified gravity are then taken into account via the following parametrization of Poisson and anisotropy equations:
\begin{align}\label{MGCAMB}
&k^2\Psi=-\mu(a,k)\frac{a^2}{2M_P^2}\left[\rho_m\Delta_m+3\left(\rho_m+P_m\right)\sigma_m\right] \,,\nonumber \\
&k^2\left[\Phi-\gamma(a,k)\Psi\right]=\mu(a,k)\frac{3a^2}{aM_P^2}\left(\rho_m+P_m\right)\sigma_m \,.
\end{align}
where no QS limit has yet been taken, and $\sigma_m$ is the anisotropic stress from matter, to which neutrinos are expected to contribute at high redshift. 

For the specific case of $f(R)$, $\mu$ and $\gamma$ are expressed in terms of a single parameter, $B_0$, as follows
\begin{align}\label{MGCAMBfR}
\mu&=\frac{1}{1-B_0 \Omega_m a^{3}/2}\frac{1+(2/3)B_0\left(k/H_0\right)^2a^4}{1+(1/2)B_0\left(k/H_0\right)^2a^4}, \nonumber \\
\gamma&=\frac{1+(1/3)B_0\left(k/H_0\right)^2a^4}{1+(2/3)B_0\left(k/H_0\right)^2a^4}
\end{align}
which correspond to a QS approximation with a power law to describe the time evolution of the Compton wavelength of the scalar d.o.f.~\cite{Hojjati:2012rf}. For the remaining, we will refer to~(\ref{MGCAMBfR}) as the \emph{BZ} parametrization.
Equations~(\ref{MGCAMBfR})  are combined with the system of Boltzmann equations for matter components and the dynamics of linear scalar perturbations is evolved. They are numerically implemented in~\cite{Zhao:2008bn,Hojjati:2011ix}.
\subsection{Results and Discussion}\label{results}
\begin{figure*}[!th]
\begin{center}
\includegraphics[width=0.84\textwidth]{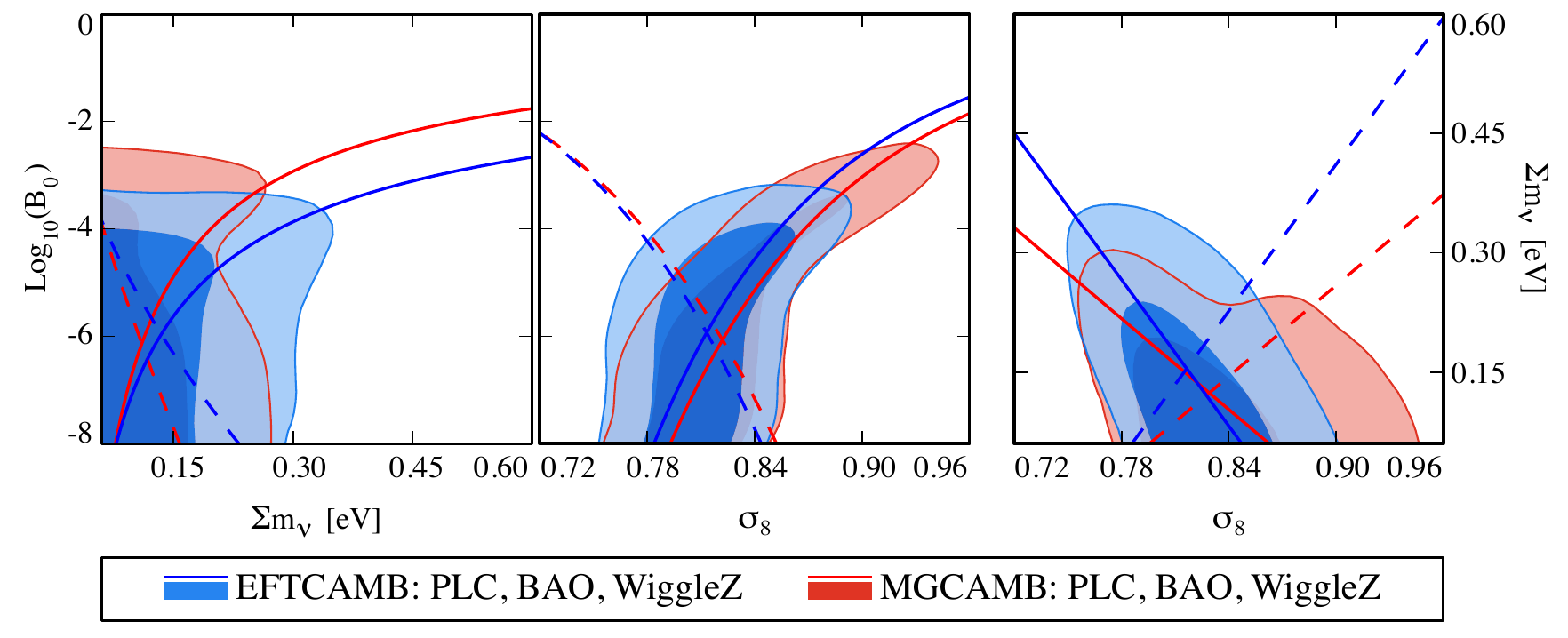}
\caption{The marginalized joint likelihood for the present day value of  ${\rm Log}_{10} B_0$, the sum of neutrino masses, $\sum m_\nu$, and  the amplitude of the (linear) power spectrum on the scale of $8\, h^{-1}{\rm Mpc}$, $\sigma_8$. Different colors correspond to the different codes used and, hence, a different modeling of $f(R)$ as shown in the legend. The darker and lighter shades correspond respectively to the $68\%$ C.L. and the $95\%$ C.L.. The solid line indicates the best constrained direction in parameter space while the dashed line indicates the worst constrained one. As we can see these directions differ noticeably for the two codes.}
\label{Fig:MGandEFT}
\end{center}
\end{figure*}
Armed with the full and consistent treatment of the background cosmology for $f(R)$ theories in the presence of massive neutrinos, we can now turn our attention to the dynamics of linear scalar perturbations. While EFTCAMB provides a number of different parametrizations for the effective dark energy equation of state that can be used in conjunction with the $f(R)$ designer approach, in this paper we will consider only the case of a $\Lambda$CDM background. 
We compare our results with those obtained with the publicly available code MGCAMB~\cite{Zhao:2008bn,Hojjati:2011ix}, and in particular with its adapted version of~\cite{Dossett:2014oia}. We will focus on the constraints on the mass of neutrinos and on the $B_0$ parameter labeling $f(R)$ theories.

Combining the different cosmological observables described in Section~\ref{data}, we explore constraints on designer $f(R)$ models in the massive neutrinos scenario, both in the case of varying and fixed neutrino mass. The results are summarized in Table~\ref{Tab:ConstraintsFR}. 
From there we can see that the PLC data set combined with BAO weakly constrain $f(R)$ models so that in the range of interest there is no statistically significant bound on ${\rm Log}_{10} B_0$ both in the case of varying and fixed neutrino mass even though the mass of neutrinos is strongly constrained. 
This is because the constraining power of CMB temperature-temperature spectrum on $f(R)$ models is dominated by the ISW effect on the large scales.  As shown in~\cite{Lombriser:2010mp,Hu:2013aqa}, the tension between the observed low value of large-scale multipoles of CMB temperature-temperature spectrum and $\Lambda$CDM prediction could indeed be reconciled by a large value of $B_0$ because of the ISW effect. On the other hand, the summed neutrino mass is constrained better by small scale data and is affected negligibly by the tension in the low-$\ell$ multipoles.
This is confirmed by the black line in panels (a) and (b) of Figure~\ref{Fig:B0Experiment} that clearly show that the posterior probability distribution of ${\rm Log}_{10} B_0$ is peaked at a very large value because of this effect and regardless of the mass of neutrinos. 

The situation changes when  CMB lensing data are added since both $f(R)$ and massive neutrinos can affect those significantly and in a degenerate way. As we already discussed, larger values of $B_0$ correspond to a larger enhancement of the growth of structure  below the Compton scale, while larger neutrino mass corresponds to a smaller free-streaming scale and hence a smaller suppression of the growth of structure below the mass scale. Hence, when lensing data are considered $B_0$ and $\sum m_\nu$ display a significant degeneracy which is noticeable in panel (a) of Figure~\ref{Fig:B0Experiment} and  the bounds on  these quantities get worse as can be seen in Table~\ref{Tab:ConstraintsFR}.
Comparing the 1D likelihoods for $B_0$, {\it i.e.} the red curves, in the middle and right panels of Figure~\ref{Fig:B0Experiment} we can see a significant difference in the bounds between the case of varying and fixed neutrino mass. The present value of the Compton scale is well constrained when the neutrino mass is kept fixed ($\sum m_{\nu}=0.06{\rm eV}$) , with a bound of ${\rm Log}_{10}B_0<-2.3$ at $95\%$ C.L.. Instead, when the neutrino mass is varied the bound is looser. 
 
Finally we can see that the addition of the WiggleZ data improves the situation. It is true that $f(R)$ and massive neutrinos leave a degenerate imprint on LSS, however the constraining power of WiggleZ, and its high sensitivity to changes in $B_0$ within the range that we consider are able to partially alleviate the degeneracy. Hence we obtain a stringent bound on $B_0$ when the mass of neutrinos is fixed and, more  generally, substantial bounds on both $B_0$ and $\sum m_\nu$ when the mass of neutrinos is varied.  

We compare our results with those obtained with MGCAMB, following the implementation described in Section~\ref{Sec:IIA}. The results are in good agreement, even though there are some interesting differences that we shall discuss in the following.
Form Table~\ref{Tab:ConstraintsFR} we can see that the constraints obtained with EFTCAMB on ${\rm Log}_{10} B_0$ are a bit tighter while the bound on neutrino mass is weaker.
The reason why this happens can be easily understood by looking at the marginalized joint likelihood of ${\rm Log}_{10} B_0$ and $\sum m_\nu$ in Figure~\ref{Fig:MGandEFT}.
From there we can see that there is a change in the degeneracy between these two parameters. This conclusion is further confirmed by looking at the principal components of the two parameters. These are shown in Figure~\ref{Fig:MGandEFT} as two lines: the continuous one corresponds to the best constrained direction in parameter space while the dashed one corresponds to the worst constrained one. The blue lines correspond to results obtained with EFTCAMB, while the red lines correspond to those obtained with MGCAMB. As we can see the principal directions for the two codes differ noticeably. 
The same conclusion can be drawn from the other two panels of Figure~\ref{Fig:MGandEFT} where we can see that the degeneracies between ${\rm Log}_{10} B_0$, $\sigma_8$ and $\sum m_\nu$ change substantially. In particular,  there is less degeneracy between $\sigma_8$ and $\sum m_\nu$ in $f(R)$ cosmologies when the analysis is performed with EFTCAMB.

These changes in the degeneracies are due to the different modeling of modified gravity physics.  As discussed in Section~\ref{Sec:IIA}, with respect to the complete $f(R)$ modeling of EFTCAMB the modeling of MGCAMB relies on two different assumptions, namely the QS  regime for perturbation and the power law ansatz for the time dependence of the Compton wavelength of the scalaron, characteristic of the BZ parametrization. Depending on the observables under consideration and on the values of $B_0$ one approximation will affect results more than the other; a more detailed and quantitative analysis  of the roles that these two assumptions play is subject of an ongoing investigation.
\begin{figure*}[!htb]
\begin{center}
\includegraphics[width=0.84\textwidth]{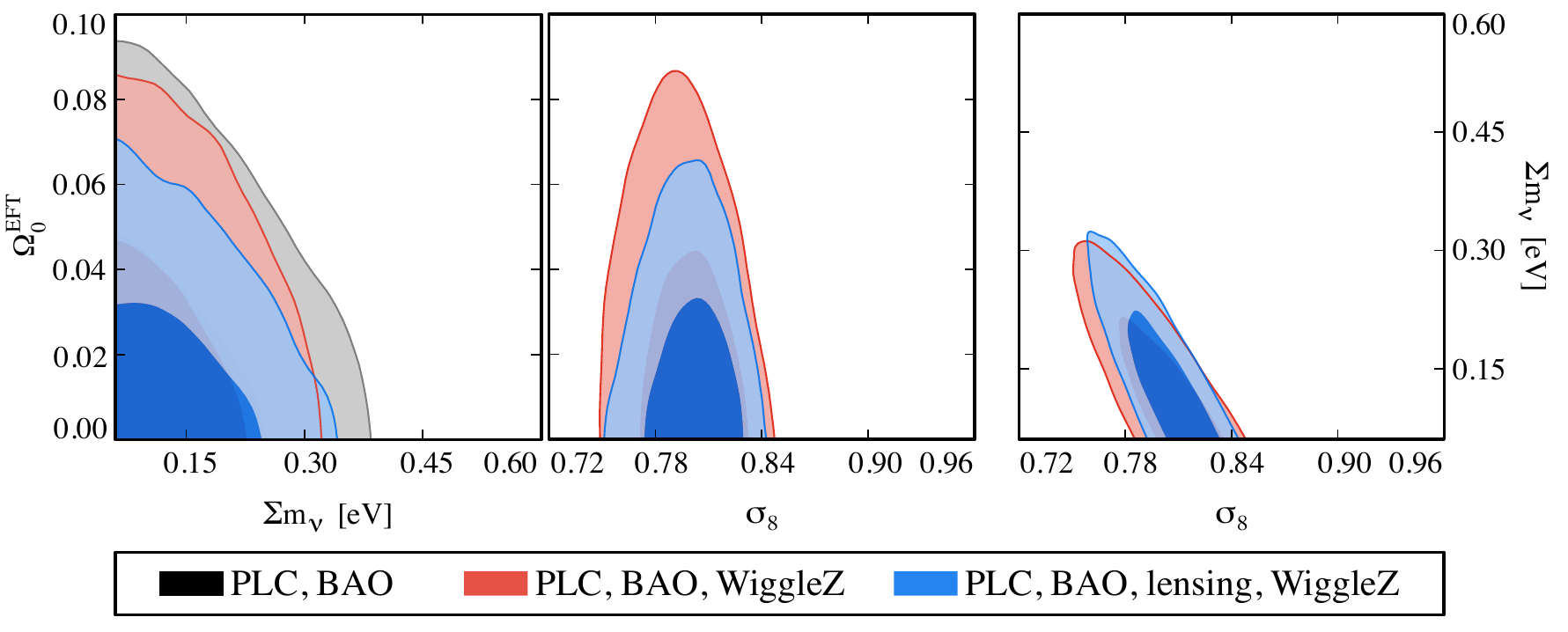}
\caption{The marginalized joint likelihood for the present day value of  $\Omega_0^{\rm EFT}$, the sum of neutrino masses, $\sum m_\nu$, and  the amplitude of the (linear) power spectrum on the scale of $8\, h^{-1}{\rm Mpc}$, $\sigma_8$. Different colors correspond to different combinations of cosmological observations as shown in the legend. The darker and lighter shades correspond respectively to the $68\%$ C.L. and the $95\%$ C.L.. No new significant degeneracies between these parameters are found.}
\label{Fig:LEFTExperiment}
\end{center}
\end{figure*}
\begin{table*}[htb!]
\footnotesize
\centering
\begin{tabular}{||l||c|c||}
\hline
\multicolumn{1}{||l||}{  }&
\multicolumn{2}{c||}{Varying $m_{\nu}$} \\
\hline
\hline

Data sets & $\Omega_0^{\rm EFT}$ $(95\% \,\,{\rm C.L.})$ & $\sum m_{\nu}$ $(95\% \,\,{\rm C.L.})$  \\
\hline
PLC + BAO 								      & $< 0.06$ & $< 0.30$  \\
PLC + BAO + WiggleZ					  & $< 0.06$ & $< 0.25$ \\
PLC + BAO + lensing + WiggleZ      & $< 0.05$ & $< 0.26$ \\
\hline
\end{tabular}
\caption{Constraints on the cosmological parameter of \textit{pure} linear EFT parametrization with varying neutrino mass, using different combinations of data sets.}
\label{Tab:ConstraintsLEFT}
\end{table*}
\section{A worked example II: massive neutrinos and pure EFT models}\label{Sec:linearEFTmassiveNu}
Another built in case that can be explored with EFTCAMB is the so called \textit{pure} EFT mode, in which one parametrizes directly the time dependence of the different functions in the EFT action~\cite{Hu:2013twa,Raveri:2014cka}.  For simplicity,  we will focus on models that contain only the three background operators, {\it i.e.} the only ones that affect also the dynamics of the background besides that of the perturbations. The corresponding action is

\begin{align}\label{EFT_action}
 S = \int d^4x &\sqrt{-g}   \left \{ \frac{m_0^2}{2} \left[1+\Omega(\tau)\right]R+ \Lambda(\tau) - a^2c(\tau) \delta g^{00}\right\}\nonumber\\
 +&S_2[g_{\mu \nu}]+ S_{m} [\chi_i ,g_{\mu \nu}],
\end{align}
where $S_2$ contains all EFT operators affecting only the dynamics of perturbations and $S_m$ is the action for matter fields, which are minimally coupled to the metric $g_{\mu\nu}$.

As discussed in~\cite{Hu:2013twa}, we employ a  designer approach and, after fixing the desired expansion history, we eliminate two of the three EFT functions in favor of the third one, which is typically chosen to be $\Omega(a)$; to fully specify the model we then need to choose a time dependence for the latter~\cite{Frusciante:2013zop}.  In this paper we investigate degeneracies between massive neutrinos and \textit{pure} EFT models with a linear parametrization of $\Omega$, {\it i.e.} 
\be
\label{PureEFT:LinearEFT}
\Omega(a) = \Omega_0^{\rm EFT} a\;.
\ee
This model can be considered as a parametrization of a scalar-tensor theory, where $\Omega$ is the coupling function and  the kinetic  and the potential terms are represented respectively by the c function and $c-\Lambda$. Following the prescription in~\cite{Hu:2013twa} once the background expansion history is chosen also the potential and the kinetic terms result to be fixed.       
While EFTCAMB allows to choose between several  expansion histories in the following analysis we fix it to match the $\Lambda$CDM one. 

To test this model we used different combinations of the data set described in Section~\ref{data}; the corresponding constraints on $\Omega_0^{\rm EFT}$ and $\sum m_{\nu}$ are listed in Table~\ref{Tab:ConstraintsLEFT}. From there we can see that the bounds on $\Omega_0^{\rm EFT}$ do not sensibly change with respect to the results reported in~\cite{Raveri:2014cka}, where neutrino masses were set to zero. In that case the bound on $\Omega_0^{\rm EFT}$ was found to be $\Omega_0^{\rm EFT}<0.061$ (95\% C.L.) when considering \textit{Planck}+WP+BAO+lensing (for details see Table I(a) in~\cite{Raveri:2014cka}).  The only exception is given by the most complete data combination which slightly improves on the previous bounds. Overall, the constraints on the sum of neutrino masses are  more stringent than in the $f(R)$ case for all the data sets considered, given the lack of degeneracy as can be noticed in Figure~\ref{Fig:LEFTExperiment}. In particular the bounds are close to those found in~\cite{Ade:2013zuv} in the absence of modified gravity.
 However, when lensing data are added one can notice some weak degeneracy between $\Omega_0^{\rm EFT}$ and $\sum m_{\nu}$ in the left panel of Figure~\ref{Fig:LEFTExperiment}, which results in this data set favoring a slightly bigger neutrino mass and a smaller value of $\Omega_0^{\rm EFT}$. From the middle and right panels of Figure~\ref{Fig:LEFTExperiment}, we can see that the degeneracy between $\sigma_8$ and $\Omega_0^{\rm EFT}$ is also negligible and that the interplay between neutrino masses and $\sigma_8$ is not sensibly altered, with respect to~\cite{Ade:2013zuv}, by linear EFT models.

In conclusion, our results suggest that no degeneracy with massive neutrinos is present when the linear EFT model on $\Lambda$CDM cosmology is considered. One could wonder whether the same model on a different background would result in a different degeneracy. The results in~\cite{Raveri:2014cka} seem to imply that exploring the same model on $w$CDM would not alter so much the bounds, since  no effects due to a modification of gravity would be appreciable. It would be certainly interesting to explore it on the CPL background~\cite{Chevallier:2000qy,Linder:2002et}.  Certainly another line to explore in order to find sizable differences from general relativity would be that of  investigating more complex temporal evolution of the non-minimally coupling constant $\Omega(a)$ than linear model.
\section{Conclusions}
It is well known that the addition of massive neutrinos to the standard cosmological model affects the growth of structures in the Universe. On the other hand, the same imprint on structure formation  might be also a characteristic of a class of scalar tensor theories. That is why the degeneracy between the massive neutrino component and models of modified gravity has been extensively investigated, in particular for  $f(R)$ theories. 

In this work we used EFTCAMB/EFTCosmoMC to investigate the degeneracy between massive neutrinos and generalized theories of gravity in a $\Lambda$CDM background. In particular we considered designer $f(R)$ models and a simple linear EFT model, effectively consisting of a scalar field whose coupling to gravity changes linearly with the scale factor.  In each case, we considered different combinations of the first released data of the  {\it Planck} satellite, BAO measurements from as well as large scale structure data from Wiggle Z.

In the $f(R)$ case, we found that the combination of  {\it Planck} and BAO measurements displayed a marked degeneracy between the Compton wavelength of the scalaron and the sum of neutrino masses.
With the addition of large scale structure data from the WiggleZ experiment we found that this degeneracy is alleviated resulting in stronger constraints.
In particular the most complete data set that we used results in ${\rm Log}_{10} B_0<-4.1$ at $95\% \,\,{\rm C.L.}$ if neutrinos are assumed to have a fixed mass equal to $\sum m_\nu = 0.06 \, {\rm eV}$ and ${\rm Log}_{10} B_0<-3.8$ and $\sum m_\nu <0.32$ at $95\% \,\,{\rm C.L.}$ if neutrino masses are allowed to vary. We compared our results to those obtained by means of MGCAMB, in which $f(R)$ models are typically treated via parametrization introduced in~\cite{Bertschinger:2008zb}, which assumes the quasi static regime and a specific power law evolution for the characteristic lengthscale of the model. Overall there is good agreement between the two codes, finding however that  due to the different modeling there is a slight change in the degeneracy between $f(R)$ models and massive neutrinos. 
In particular this degeneracy changes direction in parameter space resulting in the fact that EFTCAMB obtains stronger bounds on ${\rm Log}_{10} B_0$ but weaker constraints on $\sum m_\nu$.
We observed that also other degeneracies are affected by the different physical modeling so that with EFTCAMB there is less degeneracy between $f(R)$ models and  $\sigma_8$.

In the case of the \textit{pure} linear EFT model we found that, in contrast to the $f(R)$ case, there is no appreciable degeneracy between the present day value of the coupling, $\Omega_0^{\rm EFT}$, and the sum of neutrino masses for all the data set combinations that we considered. As a result the constraints on $\Omega_0^{\rm EFT}$ slightly improve with respect to the one previously obtained in~\cite{Raveri:2014cka} regardless of the presence of massive neutrinos. The combination of the PLC, BAO, lensing and WiggleZ data then results in $\Omega_0^{\rm EFT}<0.05$ and $\sum m_\nu <0.26$ at $95\% \,\,{\rm C.L.}$.

This work was made possible by the release of an updated version of EFTCAMB/EFTCosmoMC which is publicly available at \url{http://wwwhome.lorentz.leidenuniv.nl/~hu/codes/} and includes the full compatibility with massive neutrinos of all the built-in modified gravity models. In addition, it includes the complete modeling of the effects of modified gravity on tensor modes and polarization spectra. The designer section of the code has been expanded with the inclusion of designer minimally coupled quintessence models and the stability priors have been equipped with several options to control the ones that are not related to mathematical stability.
Finally, the updated version includes several new parametrizations for the equation of state of DE that are also fully compatible with designer models.

\begin{acknowledgments}
BH is supported by the Dutch Foundation for Fundamental Research on Matter (FOM). MR acknowledges partial support from the INFN-INDARK initiative. AS acknowledges support from the D-ITP consortium, a program of the Netherlands Organisation for Scientific Research (NWO) that is funded by the Dutch Ministry of Education, Culture and Science (OCW). NF acknowledges partial financial support from the European Research Council under the European Union's Seventh Framework Programme (FP7/2007-2013) / ERC Grant Agreement n.~306425 ``Challenging General Relativity''. MR thanks the Instituut Lorentz (Leiden University) for hospitality while this work was being completed. 
\end{acknowledgments}


\begin{thebibliography}{99}

\bibitem{Maltoni:2004ei} 
  M.~Maltoni, T.~Schwetz, M.~A.~Tortola and J.~W.~F.~Valle,
  New J.\ Phys.\  {\bf 6}, 122 (2004)
  [hep-ph/0405172].
  
\bibitem{Fogli:2005cq} 
  G.~L.~Fogli, E.~Lisi, A.~Marrone and A.~Palazzo,
  Prog.\ Part.\ Nucl.\ Phys.\  {\bf 57}, 742 (2006)
  [hep-ph/0506083].
     
\bibitem{Lesgourgues:2006nd} 
  J.~Lesgourgues and S.~Pastor,
  Phys.\ Rept.\  {\bf 429}, 307 (2006)
  [astro-ph/0603494].
  
\bibitem{Wong:2011ip} 
  Y.~Y.~Y.~Wong,
  Ann.\ Rev.\ Nucl.\ Part.\ Sci.\  {\bf 61}, 69 (2011)
  [arXiv:1111.1436 [astro-ph.CO]].

\bibitem{Ade:2013zuv} 
  P.~A.~R.~Ade {\it et al.}  [Planck Collaboration],
  Astron.\ Astrophys.\  (2014)
  [arXiv:1303.5076 [astro-ph.CO]].
  
\bibitem{Lewis:2002nc} 
  A.~Lewis and A.~Challinor,
  Phys.\ Rev.\ D {\bf 66}, 023531 (2002)
  [astro-ph/0203507].
  
\bibitem{Silvestri:2009hh} 
  A.~Silvestri and M.~Trodden,
  Rept.\ Prog.\ Phys.\  {\bf 72}, 096901 (2009)
  [arXiv:0904.0024 [astro-ph.CO]].
  
\bibitem{Jain:2010ka} 
  B.~Jain and J.~Khoury,
  Annals Phys.\  {\bf 325}, 1479 (2010)
  [arXiv:1004.3294 [astro-ph.CO]].
  
\bibitem{Jain:2013wgs} 
  B.~Jain, A.~Joyce, R.~Thompson, A.~Upadhye, J.~Battat, P.~Brax, A.~C.~Davis and C.~de Rham {\it et al.},
  arXiv:1309.5389 [astro-ph.CO].
  
\bibitem{Baldi:2013iza} 
  M.~Baldi {\it et al.}, 
  arXiv:1311.2588 [astro-ph.CO].
  
\bibitem{He:2013qha} 
  J.~h.~He,
  Phys.\ Rev.\ D {\bf 88}, no. 10, 103523 (2013)
  [arXiv:1307.4876 [astro-ph.CO]].
  
\bibitem{Dossett:2014oia} 
  J.~Dossett, B.~Hu and D.~Parkinson,
  JCAP {\bf 1403}, 046 (2014)
  [arXiv:1401.3980 [astro-ph.CO]].
  
\bibitem{Hojjati:2011ix} 
  A.~Hojjati, L.~Pogosian and G.~B.~Zhao,
  JCAP {\bf 1108}, 005 (2011)
  [arXiv:1106.4543 [astro-ph.CO]].
  
\bibitem{Motohashi:2012wc} 
  H.~Motohashi, A.~A.~Starobinsky and J.~Yokoyama,
  Phys.\ Rev.\ Lett.\  {\bf 110}, no. 12, 121302 (2013)
  [arXiv:1203.6828 [astro-ph.CO]].
    
\bibitem{Hu:2013twa} 
  B.~Hu, M.~Raveri, N.~Frusciante and A.~Silvestri,
  Phys.\ Rev.\ D {\bf 89}, 103530 (2014)
  [arXiv:1312.5742 [astro-ph.CO]].
	
\bibitem{Raveri:2014cka} 
  M.~Raveri, B.~Hu, N.~Frusciante and A.~Silvestri,
  Phys.\ Rev.\ D {\bf 90}, 043513 (2014)
  [arXiv:1405.1022 [astro-ph.CO]].
  
\bibitem{CAMB}
\url{http://camb.info} \,.

\bibitem{Lewis:1999bs} 
  A.~Lewis, A.~Challinor and A.~Lasenby,
  Astrophys.\ J.\  {\bf 538}, 473 (2000),
  [astro-ph/9911177].
  
\bibitem{Lewis:2002ah}
  A.~Lewis and S.~Bridle,
  Phys.\ Rev.\ D {\bf 66}, 103511 (2002)
  [astro-ph/0205436].
	
\bibitem{Gubitosi:2012hu} 
  G.~Gubitosi, F.~Piazza and F.~Vernizzi,
  JCAP {\bf 1302}, 032 (2013)
  [arXiv:1210.0201 [hep-th]].
  
\bibitem{Bloomfield:2012ff} 
  J.~K.~Bloomfield, \'E. \'E.~Flanagan, M.~Park and S.~Watson,
  JCAP {\bf 1308}, 010 (2013)
  [arXiv:1211.7054 [astro-ph.CO]].
  
\bibitem{Piazza:2013coa} 
  F.~Piazza and F.~Vernizzi,
  Class.\ Quant.\ Grav.\  {\bf 30}, 214007 (2013)
  [arXiv:1307.4350].
  
\bibitem{Jassal:2004ej}
 H.~K.~Jassal, J.~S.~Bagla and T.~Padmanabhan,
 Mon.\ Not.\ Roy.\ Astron.\ Soc.\  {\bf 356}, L11 (2005)
 [astro-ph/0404378].

\bibitem{Jassal:2006gf}
 H.~K.~Jassal, J.~S.~Bagla and T.~Padmanabhan,
 Mon.\ Not.\ Roy.\ Astron.\ Soc.\  {\bf 405}, 2639 (2010)
 [astro-ph/0601389].

\bibitem{Hu:2014ega}
 Y.~Hu, M.~Li, X.~-D.~Li and Z.~Zhang,
 Sci.\ China Phys.\ Mech.\ Astron.\  {\bf 57}, 1607 (2014)
 [arXiv:1401.5615 [astro-ph.CO]].
	
\bibitem{Hu:2014oga} 
  B.~Hu, M.~Raveri, N.~Frusciante and A.~Silvestri,
  arXiv:1405.3590 [astro-ph.IM].

\bibitem{Ade:2013kta} 
  P.~A.~R.~Ade {\it et al.}  [Planck Collaboration],
  arXiv:1303.5075 [astro-ph.CO].

\bibitem{Hinshaw:2012aka}
  G.~Hinshaw {\it et al.}  [WMAP Collaboration],
  Astrophys.\ J.\ Suppl.\  {\bf 208}, 19 (2013),
  [arXiv:1212.5226 [astro-ph.CO]].

\bibitem{Ade:2013tyw} 
  P.~A.~R.~Ade {\it et al.}  [Planck Collaboration],
  arXiv:1303.5077 [astro-ph.CO].

\bibitem{Beutler:2011hx} 
  F.~Beutler {\it et al.},
  Mon.\ Not.\ Roy.\ Astron.\ Soc.\  {\bf 416}, 3017 (2011),
  [arXiv:1106.3366 [astro-ph.CO]].
  
\bibitem{Percival:2009xn} 
  W.~J.~Percival {\it et al.}  [SDSS Collaboration],
  Mon.\ Not.\ Roy.\ Astron.\ Soc.\  {\bf 401}, 2148 (2010),
  [arXiv:0907.1660 [astro-ph.CO]].
  
\bibitem{Padmanabhan:2012hf} 
  N.~Padmanabhan {\it et al.},
  Mon.\ Not.\ Roy.\ Astron.\ Soc.\  {\bf 427}, no. 3, 2132 (2012),
  [arXiv:1202.0090 [astro-ph.CO]].
  
\bibitem{Anderson:2012sa} 
  L.~Anderson {\it et al.},
  Mon.\ Not.\ Roy.\ Astron.\ Soc.\  {\bf 427}, no. 4, 3435 (2013),
  [arXiv:1203.6594 [astro-ph.CO]].
 
 \bibitem{wigz}
 \url{http://smp.uq.edu.au/wigglez-data} 
  
\bibitem{Drinkwater:2009sd} 
  M.~J.~Drinkwater {\it et al.},
  Mon.\ Not.\ Roy.\ Astron.\ Soc.\  {\bf 401}, 1429 (2010)
  [arXiv:0911.4246 [astro-ph.CO]].
  
\bibitem{Parkinson:2012vd} 
  D.~Parkinson {\it et al.},
  Phys.\ Rev.\ D {\bf 86}, 103518 (2012)
  [arXiv:1210.2130 [astro-ph.CO]].
  
\bibitem{Blake:2010xz} 
  C.~Blake {\it et al.},
  Mon.\ Not.\ Roy.\ Astron.\ Soc.\  {\bf 406}, 803 (2010)
  [arXiv:1003.5721 [astro-ph.CO]].

\bibitem{Bertschinger:2008zb} 
  E.~Bertschinger and P.~Zukin,
  Phys.\ Rev.\ D {\bf 78}, 024015 (2008),
  [arXiv:0801.2431 [astro-ph]].

\bibitem{Zhao:2008bn} 
  G.~B.~Zhao, L.~Pogosian, A.~Silvestri and J.~Zylberberg,
  Phys.\ Rev.\ D {\bf 79}, 083513 (2009)
  [arXiv:0809.3791 [astro-ph]].

\bibitem{Sotiriou:2008rp} 
  T.~P.~Sotiriou and V.~Faraoni,
  Rev.\ Mod.\ Phys.\  {\bf 82}, 451 (2010)
  [arXiv:0805.1726 [gr-qc]].

\bibitem{DeFelice:2010aj} 
  A.~De Felice and S.~Tsujikawa,
  Living Rev.\ Rel.\  {\bf 13}, 3 (2010)
  [arXiv:1002.4928 [gr-qc]].
  
\bibitem{Lombriser:2014dua} 
  L.~Lombriser,
  Annalen Phys.\  {\bf 526}, 259 (2014)
  [arXiv:1403.4268 [astro-ph.CO]].

\bibitem{Starobinsky:2007hu} 
  A.~A.~Starobinsky,
  JETP Lett.\  {\bf 86}, 157 (2007),
  [arXiv:0706.2041 [astro-ph]].
  
\bibitem{Song:2006ej} 
  Y.~-S.~Song, W.~Hu and I.~Sawicki,
  Phys.\ Rev.\ D {\bf 75}, 044004 (2007),
  [astro-ph/0610532].
  
\bibitem{Bean:2006up} 
  R.~Bean {\it et al.},
  Phys.\ Rev.\ D {\bf 75}, 064020 (2007),
  [astro-ph/0611321].

\bibitem{Pogosian:2007sw} 
  L.~Pogosian and A.~Silvestri,
  Phys.\ Rev.\ D {\bf 77}, 023503 (2008)
  [Erratum-ibid.\ D {\bf 81}, 049901 (2010)]
  [arXiv:0709.0296 [astro-ph]].

\bibitem{Komatsu:2010fb} 
  E.~Komatsu {\it et al.}  [WMAP Collaboration],
  Astrophys.\ J.\ Suppl.\  {\bf 192}, 18 (2011)
  [arXiv:1001.4538 [astro-ph.CO]].
    
\bibitem{Howlett:2012mh} 
  C.~Howlett, A.~Lewis, A.~Hall and A.~Challinor,
  JCAP {\bf 1204}, 027 (2012)
  [arXiv:1201.3654 [astro-ph.CO]].

\bibitem{Mangano:2001iu} 
  G.~Mangano, G.~Miele, S.~Pastor and M.~Peloso,
  Phys.\ Lett.\ B {\bf 534}, 8 (2002)
  [astro-ph/0111408].
  
\bibitem{Mangano:2005cc} 
  G.~Mangano {\it et al.},
  Nucl.\ Phys.\ B {\bf 729}, 221 (2005)
  [hep-ph/0506164].

\bibitem{Giannantonio:2009gi} 
  T.~Giannantonio, M.~Martinelli, A.~Silvestri and A.~Melchiorri,
  JCAP {\bf 1004}, 030 (2010),
  [arXiv:0909.2045 [astro-ph.CO]].
  
\bibitem{Bean:2010zq} 
  R.~Bean and M.~Tangmatitham,
  Phys.\ Rev.\ D {\bf 81}, 083534 (2010)
  [arXiv:1002.4197 [astro-ph.CO]].

\bibitem{Hojjati:2012rf} 
  A.~Hojjati, L.~Pogosian, A.~Silvestri and S.~Talbot,
  Phys.\ Rev.\ D {\bf 86}, 123503 (2012),
 [arXiv:1210.6880 [astro-ph.CO]].

\bibitem{Lombriser:2010mp} 
  L.~Lombriser, A.~Slosar, U.~Seljak and W.~Hu,
  Phys.\ Rev.\ D {\bf 85}, 124038 (2012)
  [arXiv:1003.3009 [astro-ph.CO]].
   
\bibitem{Hu:2013aqa} 
  B.~Hu, M.~Liguori, N.~Bartolo and S.~Matarrese,
  Phys.\ Rev.\ D {\bf 88}, no. 12, 123514 (2013)
  [arXiv:1307.5276 [astro-ph.CO]].
  
\bibitem{Frusciante:2013zop} 
  N.~Frusciante, M.~Raveri and A.~Silvestri,
  JCAP {\bf 1402}, 026 (2014)
  [arXiv:1310.6026 [astro-ph.CO]].
  
\bibitem{Chevallier:2000qy} 
  M.~Chevallier and D.~Polarski,
  Int.\ J.\ Mod.\ Phys.\ D {\bf 10}, 213 (2001),
 [gr-qc/0009008].
	
\bibitem{Linder:2002et} 
  E.~V.~Linder,
  Phys.\ Rev.\ Lett.\  {\bf 90}, 091301 (2003),
  [astro-ph/0208512].
  
\end{thebibliography}
\end{document}